**Atomic scale characterization of the nucleation and growth of $SnO_2$ particles in oxidized CuSn alloys**

M. Dubey, X. Sauvage*, F. Cuvilly, S. Jouen, B. Hannoyer

*University of Rouen, Groupe de Physique des Matériaux, CNRS-UMR 6634, BP-12, 76801 Saint Etienne du Rouvray Cedex, France*

*corresponding author: xavier.sauvage@univ-rouen.fr

**Abstract**

The internal oxidation of Sn was investigated to understand the oxidation kinetics of monophase CuSn alloys. $SnO_2$ particles were characterized by analytical transmission electron microscopy. The orientation relationship between $SnO_2$ and Cu was determined with a special emphasis on the atomic scale structure of Cu/$SnO_2$ interfaces (misfit dislocations and chemical structure). Habit planes with a pure oxygen plane terminating the $SnO_2$ phase are greatly favored and large misfits promote the growth of plate shaped precipitates.

At elevated temperatures (i.e. in a range of 500°C to 800°C), the oxidation of pure copper is characterized by the formation of a double oxide layer of $Cu_2O$ and $CuO$ [1-4], in ratio that depends on the temperature [5] and pressure [6]. $Cu_2O$ being a metal-deficient oxide in which the lattice diffusion of copper ions is several orders of magnitude faster than oxygen diffusion, the rate of growth of the cuprous oxide sublayer is controlled by outward diffusion of copper cations via metal vacancies [7,8]. In the top CuO layer, there are no vacancies available for fast diffusion of reacting species and hence the growth rate remains very low as compared to $Cu_2O$ [9,10]. The presence of impurities in copper plays a very important role on the oxidation mechanisms and kinetics. Zhu *et al.* suggested that impurities could explain the discrepancies in the literature data reported for the copper oxidation kinetics [11]. Impurities impede the lateral growth of both $Cu_2O$ [11] and CuO grains [2, 9] that favors grain boundary diffusion of Cu, faster than lattice diffusion, and consequently leads to an increased oxidation rate. Alloying element in solid solution in the fcc-Cu phase can also strongly affect the oxidation kinetics. Depending on the alloying element diffusivity in both the alloy and metal oxides, various scenarios are reported. For example, in the case of oxidized Cu-Si alloys, a continuous $SiO_2$ layer appears at the $Cu/Cu_2O$ interface. This layer significantly reduces the oxidation rate which is explained by the lower diffusion rate of copper ions in $SiO_2$ than in $Cu_2O$ [12, 13]. For other alloys like Cu-Ti [14] or Cu-Ni [15], internal oxidation of the alloying element was reported. The nucleation of oxide particles inside the matrix is promoted by the oxygen diffusion [14, 16] and these particles could exhibit various shapes and sizes depending on misfits and interfacial energies [17, 18]. Their influence on the overall oxidation rate of the alloys is however not fully understood and could be rather complex. The case of CuSn alloys is of particular interest because a small amount of Sn slows down the oxidation kinetic, thus this element has a protective effect against oxidation. Internal oxidation of Sn was reported in 3 to 13wt% Sn alloys oxidized in 1 atm. $O_2$ [19] and in 8.2 wt% alloy oxidized in 1 atm. laboratory air [20]. This phenomenon is attributed to the relatively low mobility of Sn. However, depending of the concentration and of the oxidation temperature, $SnO_2$ particles are detected in an inner mixed oxide layer of $Cu_2O$ and $SnO_2$ or accumulated close to the alloy/scale interface [19-22]. The aim of the present work was to clarify the exact role of tin in the oxidation kinetics of Cu-Sn alloys with a special emphasis on the nucleation and growth of $SnO_2$ nanoparticles precipitated internally in the fcc-Cu matrix.

Two model monophase CuSn alloys were prepared by cast melting (provided by KME Corporation), containing 4.2 and 8.2 wt. % Sn respectively (Cu balance). In the following, they are referred as CuSn4 and CuSn8 respectively. Samples for oxidation experiments were cut in coupons of size ~15x10x0.8 mm and homogenized under vacuum (~5x10$^{-5}$ mbar) at 600°C for 4 hr. Then, they were mechanically polished and finally cleaned with isopropanol and dried before oxidation. Oxidations were performed under laboratory static air in a muffle furnace maintained at 600°C (±2%) during four hours. After such a treatment, samples typically exhibit external multilayered oxide scales on the surface and an internal oxidation zone as discussed in [20]. To characterize the internally oxidized area, Transmission Electron Microscopy (TEM) samples were prepared from the specimen cross-section by site specific Focused Ion Beam (FIB) lift out technique [23] in a cross-beam NVISION-40 ZEISS ®. Samples were then observed in a probe corrected JEOL® ARM200F TEM operated at 200kV. Energy Filtered images were recorded using a Gatan® Imaging Filter (GIF Quantum). High Angle Annular Dark Field (HAADF) images were recorded in the scanning mode (STEM) with a 0.1 nm probe having a convergence angle of 30mrad. The collection angle on the HAADF detector was in a range of 50 to 180 mrad.

In the internal oxidation zone, the Sn in solid solution is oxidized giving rise to a two phase mixture formed by a fcc-Cu matrix and a significant volume fraction of $SnO_2$. The number density of $SnO_2$ particle is rather low, and these particles are plate shaped as observed in the internal oxidation zone of the CuSn8 alloy (Fig. 1(a)). The thickness of these platelets is in the range of 10 to 50nm, the length up to few micrometers and the width in a range of 50 to 100nm (not shown here). There is a large distribution in the inter-plate spacing, from 100 to 500nm. Energy Filtered (EF) TEM confirms that these platelets are embedded in the Cu matrix and are rich in Sn and O (Fig. 1(a)-inset). Beside, using electron diffraction, the crystallographic structure of the internal oxide platelets was identified; it matches the expected rutile structure of $SnO_2$ [24]. A diffraction pattern was recorded in the [110] zone axis of the fcc-Cu (Fig. 1 (b)), and it is interesting to note that it also fits the [1-10] zone axis of $SnO_2$. Obviously, platelets are elongated along the Cu [-110] and the $SnO_2$ [001] directions. Internal oxides particles were also characterized in STEM. Since the average atomic number in the $SnO_2$ structure is lower than that of Cu, oxide platelets appear dark on HAADF images as shown on Fig. 2(a). The internal oxidation front is clearly exhibited on this image where platelets are growing from the top left corner down to the bottom right. The high resolution STEM-HAADF image (Fig. 2(b)) was obtained along the Cu [110] zone axis. On this image, only Cu and Sn rich atomic columns are brightly imaged. The Cu/$SnO_2$ interface (arrowed) appears edge-on and the orientation relationship (OR) between Cu and $SnO_2$ is thus given by: (002) Cu // (220) $SnO_2$ (habit planes) and [-110] Cu // [001] $SnO_2$.

It is interesting to note that the lattice mismatch along this interface, between (002) $SnO_2$ and (220) Cu planes is positive (meaning that the matrix is in tension and the precipitate in compression) and relatively large ~25%.

The microstructure in the internal oxide zone of the CuSn4 alloy looks very similar, i.e. there is a significant volume fraction of $SnO_2$ elongated plates. The diffraction pattern displayed on Fig. 3(c) shows that the OR between Cu and $SnO_2$ is similar to the OR found in the CuSn8 alloy. A platelet is imaged using high resolution HAADF STEM in the [-110] zone axis of Cu, also corresponding to the [001] zone axis of $SnO_2$ (Fig.3 (a)). The imaging is thus performed parallel to the growth direction of the plate. However, there are two different habit planes: (002) Cu for the long horizontal interface and (-2-20) Cu for the short vertical interface. For the first habit plane, the lattice mismatch between (-220) $SnO_2$ and (-2-20) Cu is very large (~31%) and positive, whilst for the second, the misfit between (220) $SnO_2$ and (002) Cu is negative and much smaller (~ -7%). These misfits are compensated by dislocations lying at the interface as underlined on Fig. 4.

It is interesting to note that these large misfits do not inhibit the nucleation of the $SnO_2$ phase but surprisingly the thinnest dimension of $SnO_2$ plates does correspond to the lowest misfit. Two Sn oxides are reported in the literature, namely SnO and $SnO_2$. The free energy of formation of $SnO_2$ is much lower than that of SnO (-396 kJ/mol vs. -197 kJ/mol at 600°C [18]), so it was rather logical to observe only the $SnO_2$ phase even if in some cases the metastable SnO phase could be obtained [25]. Beside, the low number density of particles observed by TEM clearly indicates that the nucleation rate should be rather low. This is probably due to the large misfits creating large elastic distortions. Once nucleated, $SnO_2$ particles grow in the direction of the Sn rich bulk, thus the internal oxidation kinetics is obviously mostly controlled by the growth and not the nucleation. $SnO_2$ particles are plate shaped indicating that there is also some lateral growth. Indeed, while growing, these particles transform into elongated plates where the largest habit plane (002) Cu // (220) $SnO_2$ (marked (1) in Fig. 3 (a) and (b)) combines the stronger misfits between the fcc-Cu matrix and the $SnO_2$ tetragonal phase (+25% and +31%) as demonstrated by our experimental data.

Metal-oxide interfaces have been investigated theoretically and experimentally (using High Resolution TEM) in various systems [26, 27]. It has been shown that misfit dislocations along such interfaces are not defects like bulk dislocations but an integral part of the interface structure that is determined by the interaction between misfit and bonding. Thus, in the present case, the growth of $SnO_2$ plates is also controlled by these interactions across the new interface that is created. HAADF HR-STEM images provide directly both chemical (Z contrast) and structural information. Overlapping the metal-oxide interfaces with Cu and $SnO_2$ crystal structures clearly exhibit along the two major habit planes (marked (1) and (2) on Fig. 3) a polar oxide interface terminated by a pure oxygen layer (red dots). Thus, the main difference between the two habits planes shown in the Fig. 3 is the misfit and it is rather surprising that the largest dimension of $SnO_2$ plates corresponds to the direction with the highest misfits and thus the highest elastic interaction. However, one should note that these misfits are largely positive, building up some tensile stresses in the copper matrix. Since the growth of $SnO_2$ particles requires some significant volume expansion (about 7% for CuSn4 and about 13% for CuSn8), and even if part of it is probably compensated by the outward diffusion of copper; this may favor a growth with some preferred habit planes leading to a compressive state in the $SnO_2$ particles.

In conclusion, large misfits between the tetragonal $SnO_2$ and the fcc-Cu phases give rise to a low nucleation rate. In combination with the tin gradient that builds up during the internal oxidation process, this lead to the formation of plate shaped $SnO_2$ particles. The internal oxidation kinetic is controlled by the growth of theses plates. No significant differences were observed for the two alloys considered (CuSn4 and CuSn8). HR-STEM HAADF images directly show that polar oxide interface planes terminating by a pure oxygen plane are favored. However, elastic stresses play also an important role. Indeed, the elongation of the $SnO_2$ particles is promoted along directions where misfits lead a tensile state in the matrix to compensate the volume expansion required for the internal oxidation of Sn. This feature promotes the plate shape growth of $SnO_2$.

**Acknowledgements**

Authors would like to thank the Region Haute Normandie for the financial support of this work (research grant attributed to M. Dubey). M. Dubey would also like to thank N. Masquelier for fruitful discussions.

# Figure Captions

**Figure 1:** CuSn8 oxidized at 600°C for 4 hr (a) BF-TEM image showing $SnO_2$ platelets (arrowed) within the Cu matrix ; inset: EFTEM image (Cu red, Sn blue and O green), confirming that the platelets are Sn and O rich (tin oxide). (b) Corresponding SAED pattern, spots are indexed: subscript '**m**' referring to the fcc-Cu matrix in [110] zone axis and subscript '**p**', referring to the tetragonal $SnO_2$ oxide platelets in [1-10] zone axis.

**Figure 2:** CuSn8 oxidized at 600°C for 4 hr (a) low magnification HAADF-STEM images showing several parallel $SnO_2$ plates growing in the fcc-Cu matrix. (b) High magnification HAADF-STEM filtered image showing the atomic structure of the $Cu/SnO_2$ interface in [110] fcc-Cu and [1-10] $SnO_2$ zone axes.

**Figure 3:** CuSn4 oxidized at 600°C for 4 h. (a) HAADF HR-STEM image (unfiltered) of $SnO_2$ precipitate in [001] and [-110] fcc-Cu zone axes, where Sn and Cu rich columns are imaged.(b) Zooms of selected regions marked (1) and (2) in Fig.3 (a) showing the two $Cu/SnO_2$ interfaces with the overlap of Cu and $SnO_2$ crystal structures (Cu green, Sn blue, O red). The arrows indicate the termination plane in the $SnO_2$ phase, being in both case pure O. (c) SAED pattern giving the OR between the $SnO_2$ platelet '**p**' and the fcc-Cu matrix '**m**' of Fig. 3(a).

**Figure 4**
CuSn4 alloy oxidized at 600°C for 4 h- HR-STEM HAADF unfiltered image showing the two habit planes between Cu and $SnO_2$ particles. Misfit dislocations are indicated by T-shaped symbols in the $SnO_2$ phase and by arrows in the Cu matrix. Interfaces and growth ledges are highlighted by the dashed lines.

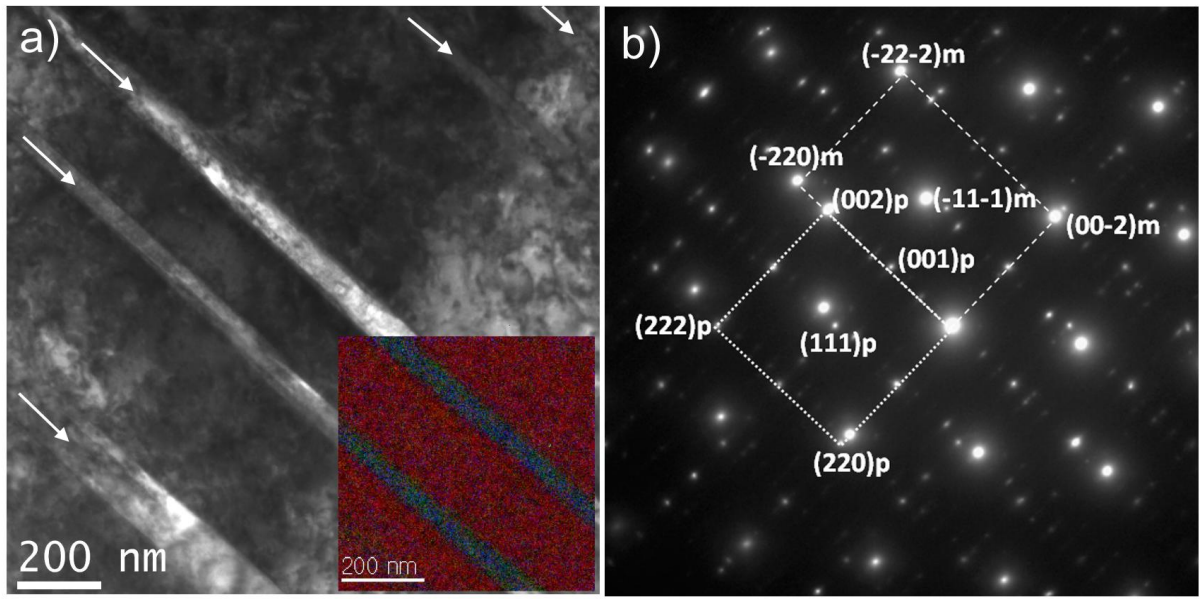

**Figure 1**

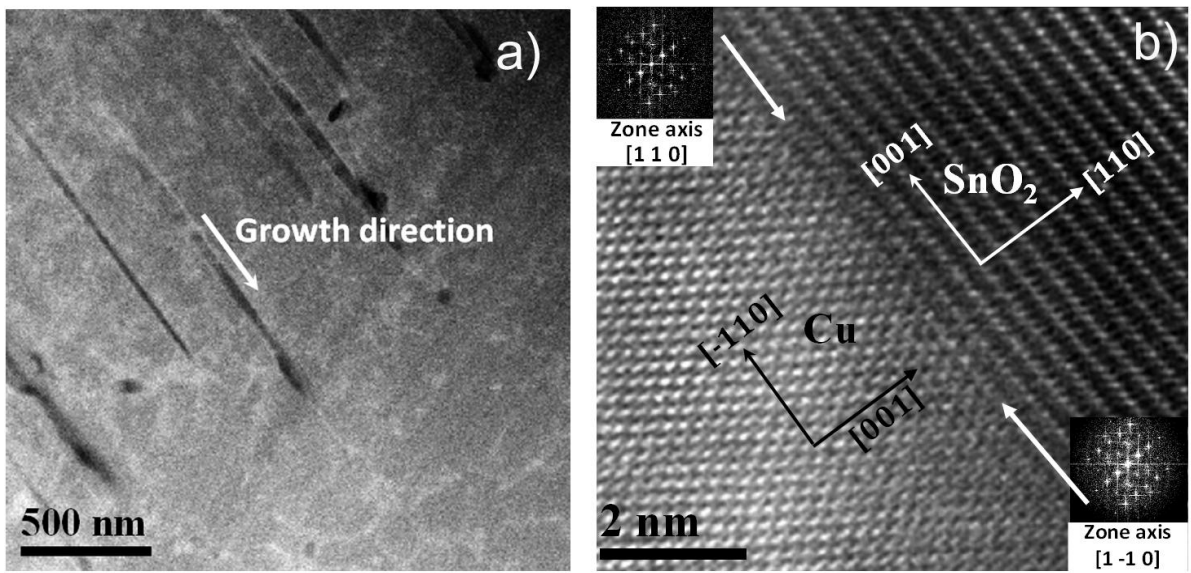

**Figure 2**

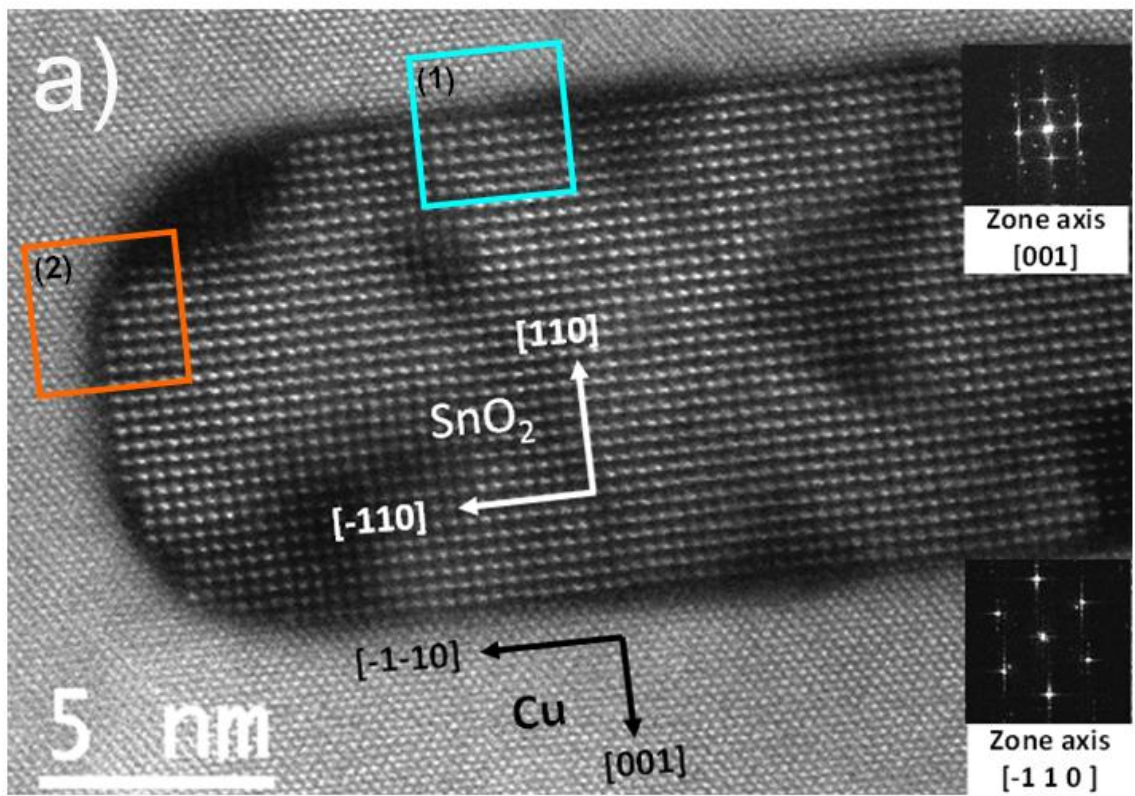
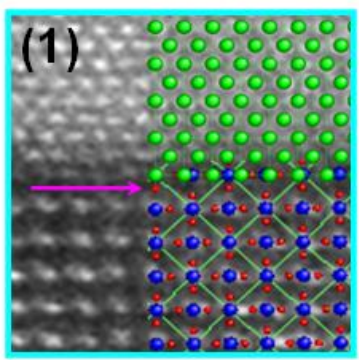
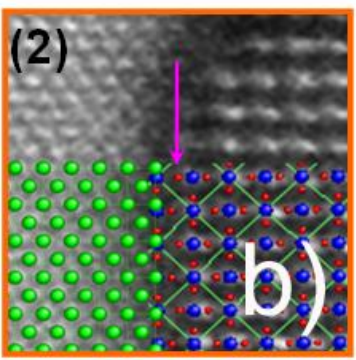
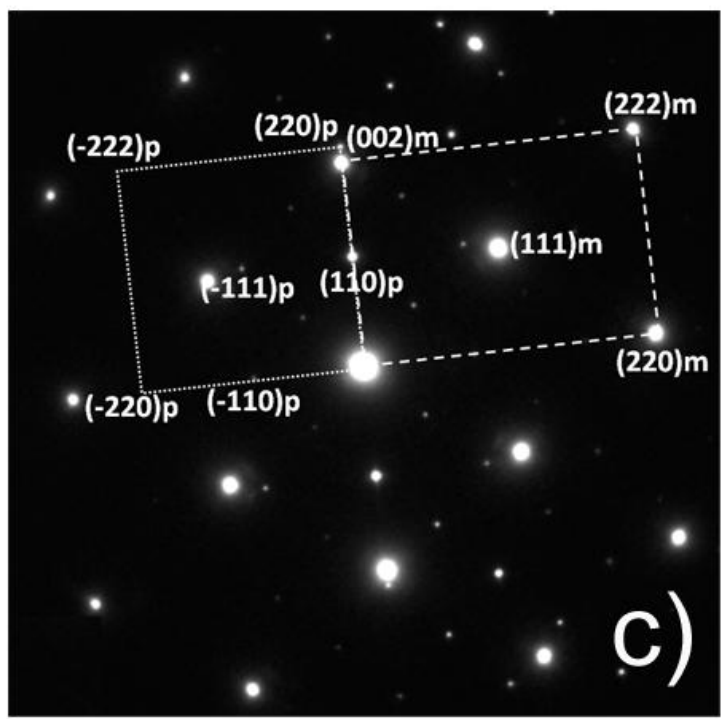

**Figure 3**

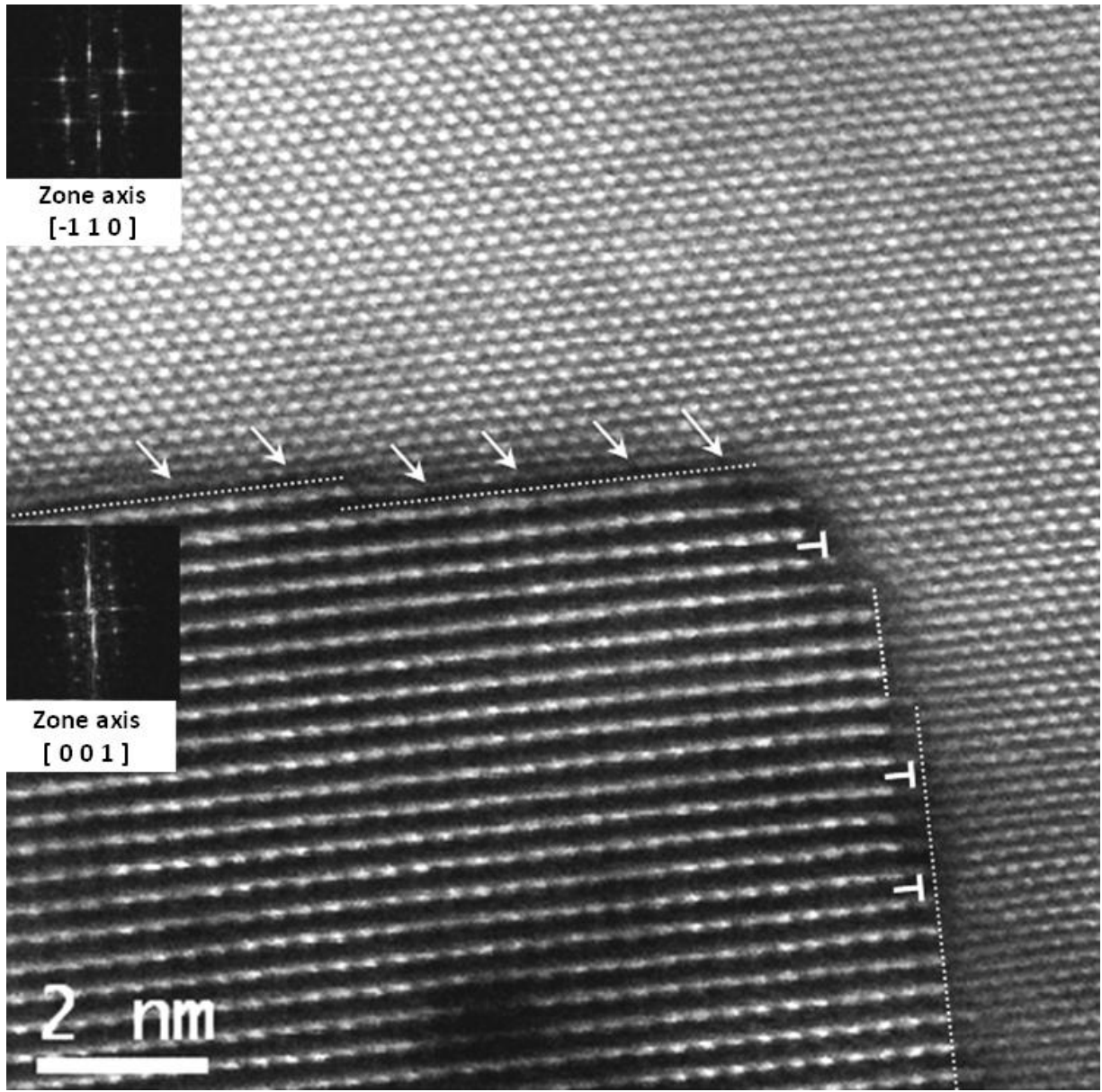

**Figure 4**